\documentclass[fleqn,usenatbib]{mnras}
\usepackage[T1]{fontenc}
\usepackage{ae,aecompl}

\usepackage{graphicx}	% Including figure files
\usepackage{amsmath}	% Advanced maths commands
\usepackage{amssymb}	% Extra maths symbols
\usepackage{times, txfonts}

\usepackage{deluxetable}

\defcitealias{2011ApJ...743...76S}{S11}
\defcitealias{2016ApJ...816...49S}{S16}
\defcitealias{2012ApJ...759..146M}{M12}
\defcitealias{2014A&A...566A..37G}{G14}

\title[CCHP: Mid--IR Colours and Metallicities of Cepheids]{The Carnegie Chicago Hubble Program: The Mid--Infrared Colours of Cepheids and the Effect of Metallicity on the CO Band--head at 4.6~$\mu$m}

\author[V. Scowcroft et al.]{Victoria~Scowcroft,$^{1}$\thanks{E-mail: vs@obs.carnegiescience.edu (VS)}
Mark Seibert,$^{1}$
Wendy~L.~Freedman,$^{2}$
Rachael~L.~Beaton,$^{1}$
\newauthor
Barry~F.~Madore,$^{1}$
Andrew~J.~Monson,$^{3}$
Jeffrey~A.~Rich,$^{1}$
Jane~R.~Rigby$^{4}$
\\
$^{1}$ Observatories of the Carnegie Institution of Washington, 813 Santa Barbara St., Pasadena, CA~91101, USA\\
$^{2}$ Department of Astronomy and Astrophysics, University of Chicago, 5640 S Ellis Ave, Chicago, IL 60637, USA \\
$^{3}$Department of Astronomy and Astrophysics, The Pennsylvania State University, 403 Davey Lab, University Park, PA, 16802, USA \\
$^{4}$Observational Cosmology Lab, NASA Goddard Space Flight Center, Greenbelt MD 20771, USA \\
}

\date{Accepted XXX. Received YYY; in original form ZZZ}

\pubyear{2016}

\begin{document}
\label{firstpage}
\pagerange{\pageref{firstpage}--\pageref{lastpage}}
\maketitle
% * <vickyscowcroft@gmail.com> 2015-09-16T21:46:28.940Z:
\begin{abstract}
We compare mid--infrared 3.6 and 4.5~$\mu$m Warm {\it Spitzer} observations for Cepheids in the Milky Way and the Large and Small Magellanic Clouds. Using models, we explore in detail the effect of the CO rotation--vibration band--head at 4.6~$\mu$m on the mid--infrared photometry. We confirm the temperature sensitivity of the CO band--head  at 4.6~$\mu$m and find no evidence for an effect at 3.6~$\mu$m. We compare the $([3.6]-[4.5])$ period--colour relations in the MW, LMC, and SMC. The slopes of the period--colour relations for the three galaxies are in good agreement, but there is a trend in zero--point with metallicity, with the lowest metallicity Cepheids having redder mid--IR colours. Finally, we present a colour--[Fe/H] relation based on published spectroscopic metallicities. This empirical relation, calibrated to the metallicity system of \citet{2014A&A...566A..37G}, demonstrates that the $([3.6]-[4.5])$ colour provides a reliable metallicity indicator for Cepheids, with a precision comparable to current spectroscopic determinations.
\end{abstract}


%%%%%%%%%%%%%%%%%%%%%%%%%%%%%%%%%%%%%%%%%%%%%%%%%%

%%%%%%%%%%%%%%%%% BODY OF PAPER %%%%%%%%%%%%%%%%%%

\section{Introduction}
\label{sec:intro}

The mid--infrared has many well-known advantages when applied to distance-scale measurements for Cepheids. First, at these long wavelengths, the effects of extinction are dramatically reduced compared to optical bands \citep[as much as 15 to 20 times smaller in the mid--infrared,][]{2005ApJ...619..931I}. Second, the intrinsic width of the mid--infrared period--luminosity (PL) relation 
is also reduced, with the dispersion of the LMC PL relation at 3.6~$\mu$m being only $\pm$0.10~mag, compared to $\pm$0.20~mag at the $V$ band \citep{2011ApJ...743...76S, 2007A&A...476...73F, 2012ApJ...745..104N}. Finally,  the mid--infrared bands are located in the Rayleigh--Jeans tail of the stellar spectrum; thus, the effect of temperature changes are minimised, such that the mid--infrared light curves are both lower amplitude (compared to their optical counterparts) and highly symmetric. 
In the optical many tens of observations are needed to sufficiently sample the `sawtooth' light curve shape. In the infrared, however, only a handful of observations are required \citep[][hereafter S11]{2011ApJ...743...76S}.
%The low amplitude means that with only a few mid--infrared observations, as compared to at least a dozen that are required in the optical, it is much more efficient to obtain the mean--light magnitude and colour of a Cepheid of known period at these longer wavelengths. 

The {\it Spitzer} Carnegie Chicago Hubble Program \citep[CCHP, P.I. W. Freedman, PID 60010,][]{2011AJ....142..192F} was designed to take advantage of these facts, employing mid--infrared 3.6~$\mu$m observations of Cepheids for its recalibration of the Hubble constant ($H_{0}$) in the mid--infrared \citep{2012ApJ...758...24F}. The CCHP has obtained observations at 3.6 and 4.5~$\mu$m from Warm {\it Spitzer}, and in the first instance expected to use both of these bands to determine distances to Cepheids in the Milky Way (MW), Large Magellanic Cloud (LMC), Local Group and beyond. However, as demonstrated by \citet{2010ApJ...709..120M}, \citetalias{2011ApJ...743...76S}, and \citet[][hereafter M12]{2012ApJ...759..146M}, the 4.5~$\mu$m band is affected by the carbon monoxide (CO) band--head at 4.6~$\mu$m. As we will show in Sections~\ref{sec:co_bandhead} and \ref{sec:pc_relation}, the abundance of CO in a Cepheid's atmosphere is strongly affected by temperature, which induces opacity variations throughout the star's pulsation cycle. This temperature dependent opacity change is echoed on a population--wide scale, varying as a function of pulsation period. The width of the 4.5~$\mu$m Cepheid period--luminosity (PL) relation is intrinsically slightly larger than the 3.6~$\mu$m PL (1.112 mag compared to 1.105 mag) due to this change in opacity. Uncorrected for the metallicity effect, the [4.5] PL relation has so far been excluded from CCHP Cepheid distance determinations \citep[see][for examples of the 4.5~$\mu$m distance modulus disagreeing with other distance moduli]{2013ApJ...773..106S, 2014ApJ...794..107R, 2016ApJ...816...49S}. The 3.6~$\mu$m Cepheid observations, however, have provided measurements of individual Cepheids precise to 4.7\% in distance, delineating the internal structure of the SMC \citep[][hereafter S16]{2016ApJ...816...49S}.

One of the outstanding problems in the study of Cepheids as distance indicators is the effect of metallicity. There has been a great deal of debate over the past few decades over the direction and size of the effect of metallicity on Cepheid magnitudes, particularly at optical wavelengths, with theoretical and empirical results often in conflict \citep[e.g.][]{2010ApJ...713..615M, 2013A&A...550A..70G}. \citet{2008A&A...488..731R} give a  review of recent measurements of the metallicity parameter $\gamma = \delta \mu / \delta \log Z$. Their summary demonstrates that a consensus is yet to be reached, with measurements ranging from $\gamma = -0.88$~mag~dex$^{-1}$ to $\gamma = +0.05$~mag~dex$^{-1}$. While most empirical measurements of $\gamma$ tend to find negative values, meaning that the true distance modulus is larger than the one obtained when neglecting the effect of the metallicity, many theoretical estimates of the parameter find the opposite. \citet{1999ApJ...522..250S} found that the effect of metallicity on the Cepheid distance modulus changed with wavelength, increasing from $\gamma = -0.02$~mag~dex$^{-1}$ at the $B$ band, to $\gamma = +0.08$~mag~dex$^{-1}$ at $V$ and to $\gamma = 0.10$ ~mag~dex$^{-1}$ in the $I$ band. \citet{1999A&A...344..551A} found a different effect in their theory--based study, concluding instead that the PL {\it slope} changed with metallicity.

More recently, \citet{2010ApJ...715..277B} studied the reddening--free Wesenheit indices constructed from optical and near--IR bands using pulsation models, finding that metallicity has a negligible effect on the $\log P - W_{VI}$ and $\log P - W_{JK}$ relations, but did significantly affect the $\log P - W_{BV}$ relation. \citet{2005ApJ...632..590M} resolved some of theoretical-empirical discrepancy by including in their models not only the dependence on the metal abundance but also the effect of the assumed  helium to metal enrichment ratio, finding a metallicity correction trend in agreement with the one based on the spectroscopic study by \citet{2005A&A...429L..37R}. Particularly notable is the study of \citet{2011ApJ...733..124S}, who found a large effect with $\gamma=-0.80$~mag~dex$^{-1}$ using Cepheids in M101. However, \citet{2011ApJ...741L..36M} demonstrate that a metallicity effect this large would bring the Cepheid distance of the Magellanic Clouds into severe disagreement with other distance indicators.

With conflicting results from both theory and observation, the CCHP chose to deal with the effect of metallicity by obtaining data at the mid--infrared wavelengths probed by {\it Spitzer}/IRAC, where the effects of metallicity were predicted to be smaller.  We will show in this paper, consistent with the results of \citet{2010ApJ...709..120M}, \citet{2011ApJ...743...76S}, and \citet{2012ApJ...759..146M}, that this is true for the observations at 3.6~$\mu$m, but not for the 4.5~$\mu$m data. However, as we have one metallicity--free band, one metallicity--dependent band, and a sub--sample of Cepheids with spectroscopic metallicities compiled by \citet[][hereafter G14]{2014A&A...566A..37G}, we can in fact {\it calibrate} the metallicity effect at 4.5~$\mu$m for the first time. 

The paper is set out as follows: In Section~\ref{sec:data} we describe the data sources we use for our analysis. In Section~\ref{sec:co_bandhead} we discuss the CO band--head present in the IRAC 4.5~$\mu$m filter, and the chemical processes that are occurring to change the opacity at this wavelength. Section~\ref{sec:pc_relation} describes the mid--infrared period--colour (PC) relations for Cepheids. In Sections~\ref{sec:metallicity_effect} and \ref{sec:metallicity_indicator} we assess the effect of metallicity on the mean PC relations, and show that the mid--infrared Cepheid colour can be used as a robust indicator of metallicity. We present our conclusions in Section~\ref{sec:conclusions}.

\section{Data}
\label{sec:data}

In this work we bring together the mid--infrared, Warm {\it Spitzer} observations taken by the CCHP \citepalias{2012ApJ...759..146M, 2011ApJ...743...76S, 2016ApJ...816...49S}, with spectroscopic metallicities compiled in \citetalias{2014A&A...566A..37G}. Many earlier studies of the effect of metallicity on Cepheid magnitudes relied on nearby H{\sc ii} region abundances as proxies for the average values of metallicities of Cepheid populations \citep{1998ApJ...498..181K, 2006ApJ...652.1133M, 2009MNRAS.396.1287S}. This technique is problematic for two reasons. First, large numbers of H{\sc ii} regions are required to trace galactic metallicity gradients. For example, in their study of M33, \citet{2008ApJ...675.1213R} found that the uncertainties in the metallicity gradients are systematically underestimated. Such underestimation would lead to incorrect inferences about the metallicity gradient, hence incorrect derivations of $\gamma$. Second, the traditional method of deriving metallicities in H{\sc ii} regions uses collisionally excited lines (CEL). These metallicities have been shown to be poor tracers of stellar abundance \citep[e.g.][]{2004cmpe.conf..115S, 2011A&A...526A..48S, 2015ApJ...798...99B}. The metallicity measurements in \citet{1994ApJ...420...87Z}, which have been adopted as standard in many Cepheid studies, use the CEL technique and as such do not trace the individual Cepheid metallicities in a meaningful way.

There are, however, individual metallicities available for our program stars that are better to use for the purposes of our study. We combine these with high--precision mid--infrared photometry to study the mid--infrared period--colour relation and its response to changing metallicity.

The data sources are described below in Sections~\ref{sec:spitzer_obs} and \ref{sec:uves_obs}. Our sample is summarised in Table~\ref{tab:data_table}; the first 10 rows are shown for information regarding its style and content. The full table is provided as supplementary material. 

\begin{table}
	\begin{center}
	\caption{Mid--infrared colours and metallicities for MW, LMC and SMC Cepheids.} 
	\label{tab:data_table}
	\begin{tabular}{l|r|r|r|r} 
		\hline \hline
Cepheid & $\log P$ & $([3.6]-[4.5])$\tablenotemark{a} & [Fe/H]\tablenotemark{b} & Galaxy \\
& (days) & \multicolumn{1}{|c|}{(mag)} & (dex) & \\
\hline
$\eta$ Aql & 0.8560 & $-0.007 \pm 0.004$ & $0.14$ & MW \\ 
U Aql & 0.8470 & $-0.015 \pm 0.004$ & $0.14$ & MW \\ 
FF Aql & 0.6500 & $0.016 \pm 0.001$ & $0.10$ & MW \\ 
SZ Aql & 1.2340 & $-0.076 \pm 0.007$ & $0.24$ & MW \\ 
TT Aql & 1.1380 & $-0.055 \pm 0.006$ & $0.19$ & MW \\ 
RT Aur & 0.5710 & $0.001 \pm 0.004$ & $0.10$ & MW \\ 
$\ell$ Car & 1.5510 & $-0.129 \pm 0.006$ & $0.24$ & MW \\ 
U Car & 1.5890 & $-0.071 \pm 0.008$ & $0.25$ & MW \\ 
CEa Cas & 0.7110 & $-0.010 \pm 0.015$ & ... &  MW \\ 
CEb Cas & 0.6510 & $0.012 \pm 0.015$ & ... &  MW \\ 
\hline
\end{tabular}
\end{center}
\tablenotetext{a}{Photometric data taken from \citetalias{2012ApJ...759..146M} (MW), \citetalias{2011ApJ...743...76S} (LMC), \citetalias{2016ApJ...816...49S} (SMC)}
\tablenotetext{b}{Spectroscopic [Fe/H] abundances taken from \citetalias{2014A&A...566A..37G} and have typical uncertainties of $\sim 0.1$~dex.}
\tablenotetext{c}{Full table available in online journal.}
\end{table}

%This paper does not need an observations section as its all pulled from other papers.
%\begin{itemize}
%\item Spiter data 
%\item Metallicity data
%\end{itemize}
\subsection{Photometric Data --- {\it Spitzer} IRAC}
\label{sec:spitzer_obs}
\begin{figure*}
\begin{center}$
\begin{array}{ccc} 
\includegraphics[width=50mm]{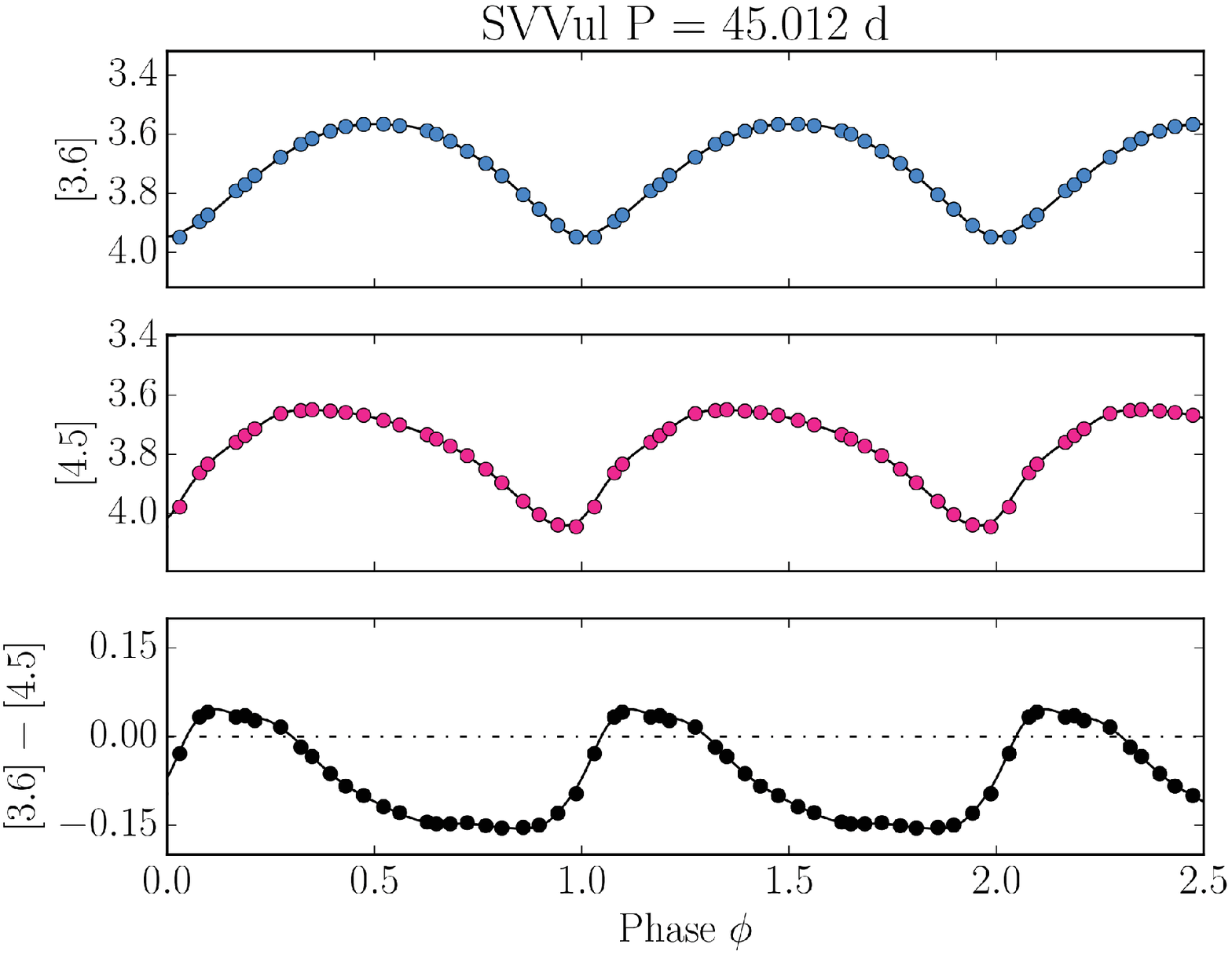} & 
\includegraphics[width=50mm]{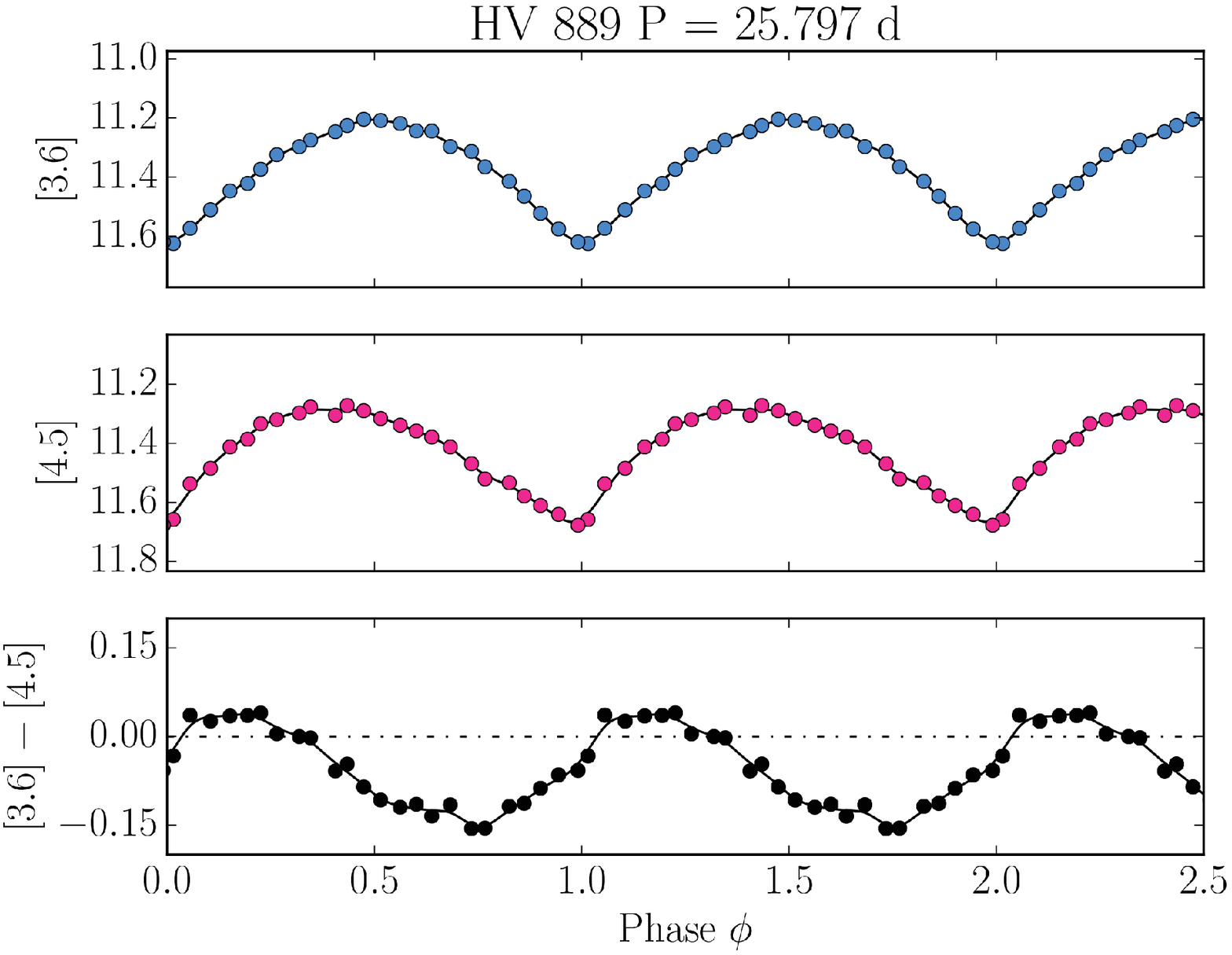} &
\includegraphics[width=50mm]{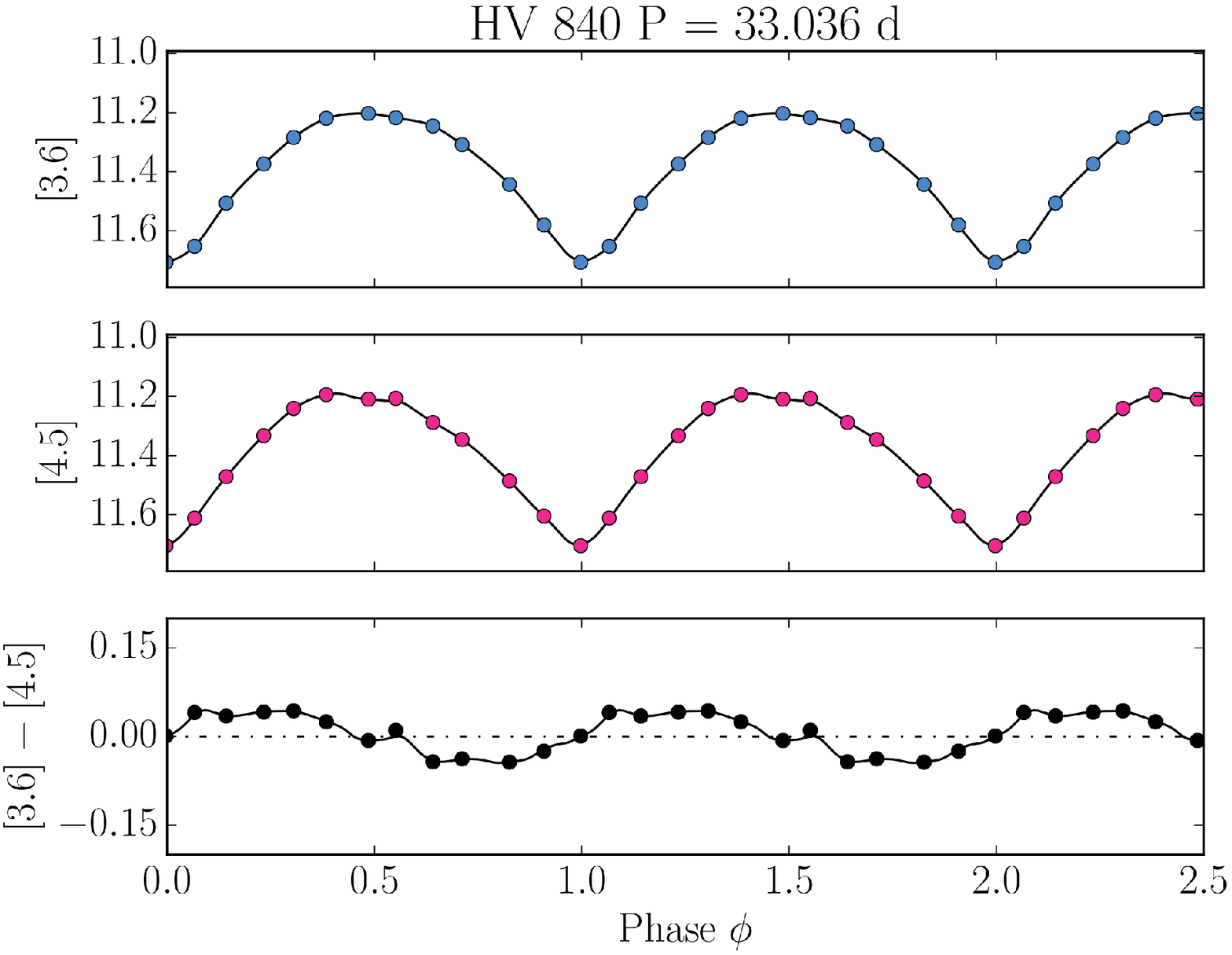} \\
\includegraphics[width=50mm]{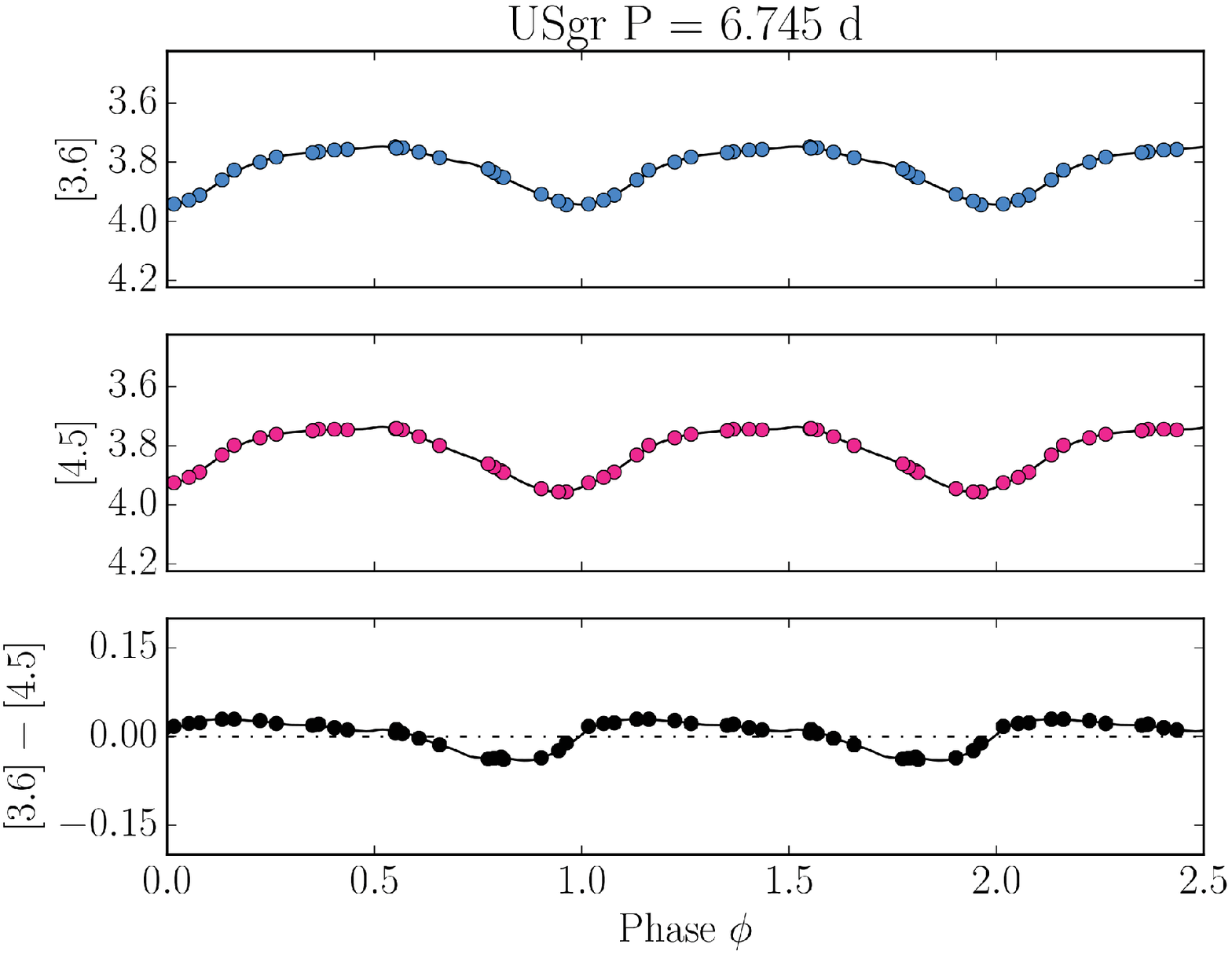} &
\includegraphics[width=50mm]{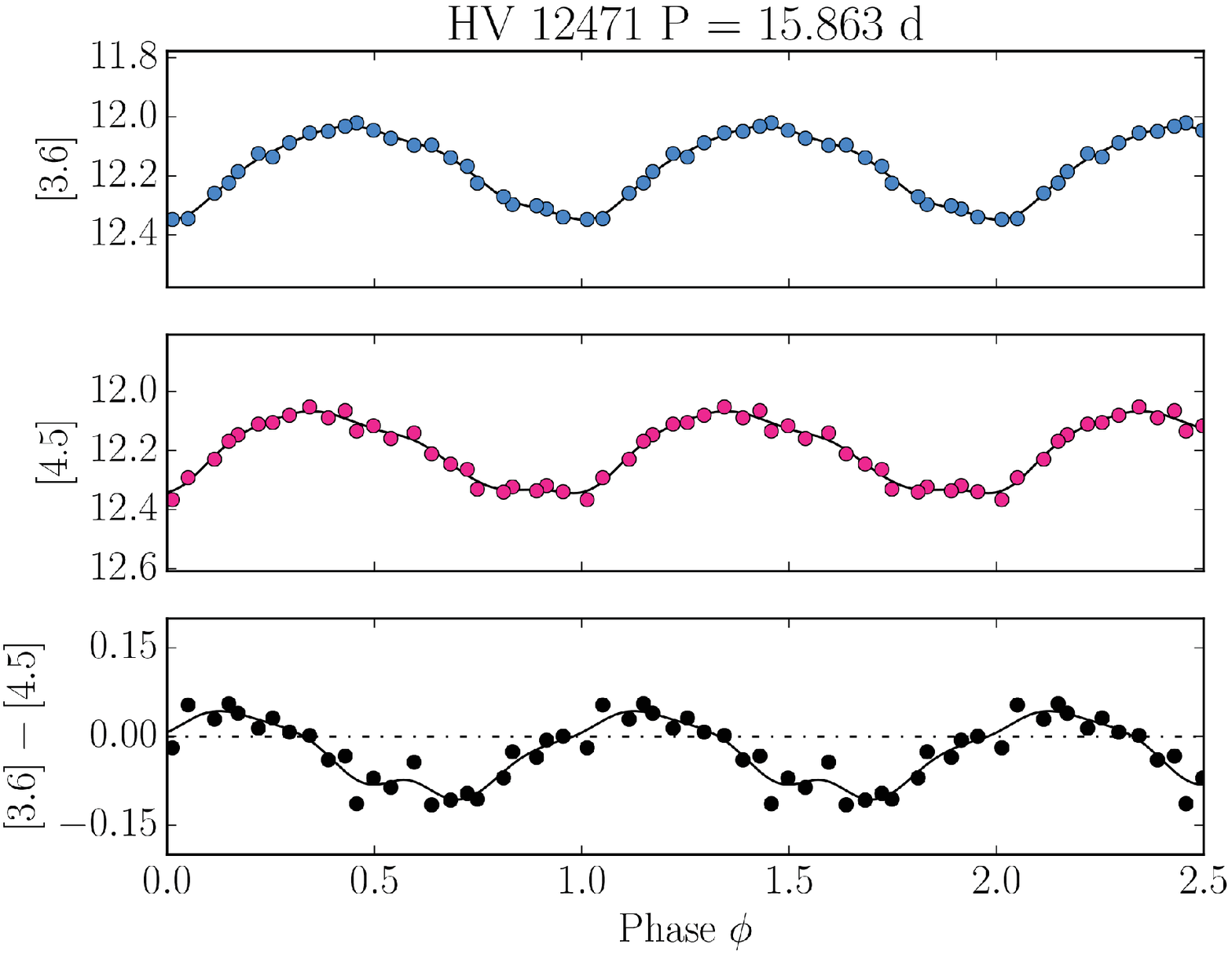} &
\includegraphics[width=50mm]{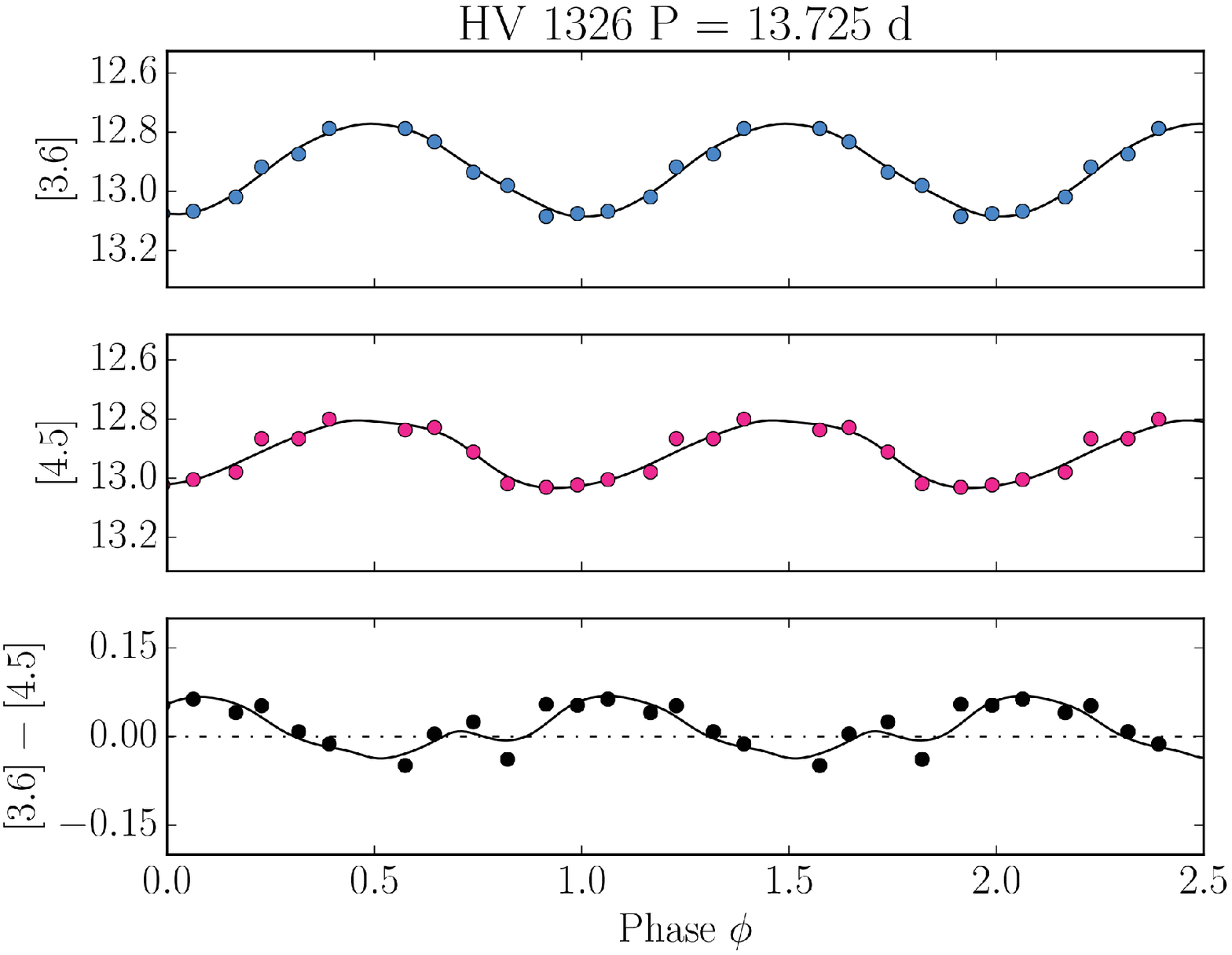} \\
\end{array}$
%% script to make these figures: plain_lightcurve.py
\caption{A selection of six Cepheid light and colour curves from Spitzer, demonstrating the data quality for the MW (left column, SV~Vul, U~Sgr), LMC (middle column, HV~899, HV~12471), and SMC (right column, HV~840, HV~1326). The three panels in each plot (from top to bottom) show the [3.6] and [4.5] light curves and the $([3.6] - [4.5])$ colour curve. The uncertainties on individual points are comparable to the size of the plotted points in the light curves. The solid lines show the {\sc gloess} fits to the light curves. Data in this figure come from \citetalias{2012ApJ...759..146M} (MW), \citetalias{2011ApJ...743...76S} (LMC), and \citetalias{2016ApJ...816...49S} (SMC).}
\label{fig:example_lcs}
\end{center}
\end{figure*}

The mid--IR data used in this work are from 3.6 and 4.5~$\mu$m Warm {\it Spitzer} IRAC observations, taken as part of the CCHP. In this work we use the photometry of \citetalias{2012ApJ...759..146M} for 37 Milky Way Cepheids, \citetalias{2011ApJ...743...76S} for 85 LMC Cepheids, and \citetalias{2016ApJ...816...49S} for 92 SMC Cepheids. Representative mid--IR light curves from the CCHP sample are shown in Figure~\ref{fig:example_lcs}. Each Cepheid is represented by three panels, with the top two showing the [3.6] and [4.5] data and {\sc gloess} \citep{2004AJ....128.2239P} fitted light curve, and the bottom panel showing the $([3.6]-[4.5])$ colour curve.

The {\it Spitzer} data were taken such that each Cepheid was observed at 24 (MW and LMC) or 12 (SMC) epochs, designed to be approximately equally spaced through a single pulsation cycle. For Cepheids with $P\leq 12$~days the observations were randomly taken, as equally spaced observations over such a short time frame would be problematic to schedule. This observing strategy gave full coverage of the light and colour variations of the Cepheid through its pulsation cycle.

The Cepheids in this sample were chosen to be useful for the CCHP distance scale calibration program. To that end, they are all isolated (i.e photometrically uncrowded), well studied stars. The MW sample was chosen to maximise overlap with high precision independent distance measurements, such as the parallax sample from \citet{2007AJ....133.1810B}. The LMC sample is focussed on longer period ($P\geq6$~days) Cepheids as these are the most readily observed in more distant galaxies. The SMC observations were taken as part of a CCHP follow up program \citep[P.I. B. Madore, PID 70010,][]{2010sptz.prop70010M}, choosing the most isolated, long period targets from the sample of previously identified Cepheids in that galaxy.

\subsection{Spectroscopic Data --- The Genovali et al. Compilation}
\label{sec:uves_obs}

We adopt metallicities from the tabulation of \citetalias{2014A&A...566A..37G}, which is the most comprehensive list of Cepheid metallicities derived from high--resolution spectroscopy for the Milky Way and Magellanic System. To build this compilation, \citetalias{2014A&A...566A..37G} combined numerous literature studies, many of which use similar instruments and/or similar methods. \citetalias{2014A&A...566A..37G} put all of the studies onto the same metallicity system by measuring zero--point offsets between different studies (see their section 4.1). While this is not a truly homogeneous sample (i.e. one taken with the same instrument and analysed identically), the \citetalias{2014A&A...566A..37G} compilation is the most complete and most homogenised sample available. Thus, it permits the largest sample of Cepheids by which to study the effects of metallicity on a star--by--star basis within the goals of this work. We refer the reader to \citetalias{2014A&A...566A..37G} (tables 2 and 4) and \citet{2013A&A...554A.132G} (table 2) for the detailed measurements for individual Cepheids in the MW (including the study originating the measurement). The Magellanic Cloud data is largely from \citet{2008A&A...488..731R}, but we refer the reader to \citetalias{2014A&A...566A..37G} (table 7) for the original data on individual stars. Typical uncertainties on the [Fe/H] abundances are approximately 0.1~dex. 

\section{The CO band--head}
\label{sec:co_bandhead}

The presence of a strong CO vibration-rotation band--head situated at 4.6~$\mu$m in stars has been known for many years \citep[for the first discussions of the ``COmosphere in late--type stars'', see][]{1994ASPC...57..124A, 1994ApJ...423..806W}. The first hints of its effects on the mid--infrared magnitudes and colours of Cepheids were noted by \citet{2010ApJ...709..120M}, who saw a large dispersion in the $([3.6]~-~[4.5])$ period--colour relation (their figure~5). They also showed theoretical models, demonstrating that as a Cepheid pulsates, (hence changes temperature) the depth of the CO feature --- squarely situated in the IRAC [4.5] band --- would also change (their figure~6). However, the \citeauthor{2010ApJ...709..120M} study was limited to single--phase observations of 29 Galactic Cepheids with a period range of $\sim 2$ to $\sim 42$~days. With the advent of Warm {\it Spitzer} and the full--phase CCHP Cepheid observations (examples of which are given in Fig.~\ref{fig:example_lcs}) we are now able to investigate the time resolved effect.

\begin{figure}
\begin{center}
\includegraphics[width=80mm]{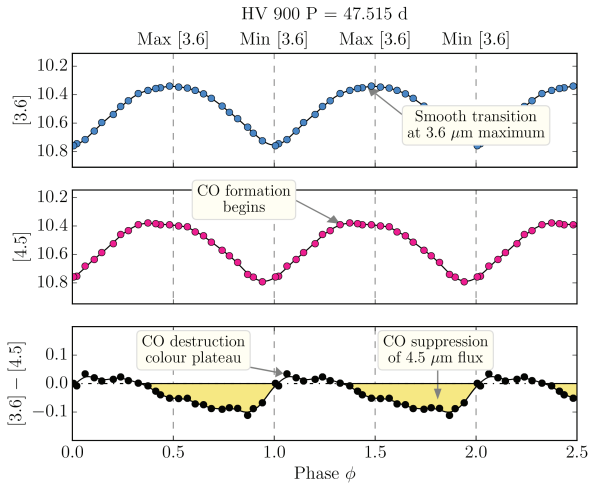}
%code annotated_lightcurve.py
\caption{The different stages of the destruction and formation of CO through the pulsation cycle of the 47.5~day LMC Cepheid HV~900. Filled points are {\it Spitzer} photometric data from \citetalias{2011ApJ...743...76S}, with error bars smaller than the size of the plotted points. Solid lines are the {\sc gloess} light curve fits to the data. The dashed vertical lines show the phase of the minimum and maximum light of the Cepheid, defined by differentiation of the [3.6] light curve. There is a slight phase lag between the min$_{[4.5]}$ and min$_{[3.6]}$. At maximum light, the [3.6] light curve flows smoothly, whereas the [4.5] light curve experiences a plateau. This plateau is due to the suppression of 4.5~$\mu$m flux by the CO band--head ``turning on'', as is seen in the bottom panel. The shaded region shows where the 4.5~$\mu$m flux is being suppressed, due to the interaction between the 4.5~$\mu$m photons and the CO molecules in the Cepheid's atmosphere. As the star reaches maximum light (hence approaches minimum radius and maximum temperature) the CO is chemically destroyed and the 4.5~$\mu$m photons are once again free to stream out of the star.}
\label{fig:annotated_lightcurve}
\end{center}
\end{figure}

\subsection{Mechanism for opacity changes at 4.5~$\mu$m}
\label{sec:mechanism}
Alongside H$_2$, CO is one of the most abundant stellar molecules, with a high dissociation energy at 11.09~eV  \citep{1984ApJS...56..193S}. 
CO is formed in stellar atmospheres by radiative association, via the reaction
\begin{equation}
\label{eqn:co_formation}
\text{C} +\text{O} \rightarrow \text{CO} + \gamma	
\end{equation}
where $\gamma$ is an emitted photon. 

Destruction by photodissociation is not prevalent in stellar atmospheres as self--shielding prevents the UV photons necessary for dissociation from reaching the molecules. For the temperatures of Cepheids, typically between 4000~K and 6000~K, the main destruction mechanism for CO is chemical dissociation, specifically reactions involving the rearrangement of atoms in neutral molecules \citep{2005hris.conf..260T}. The most important reaction at these temperatures is the endothermic reaction
\begin{equation}
\label{eqn:h_co}
\text{H} + \text{CO} \rightarrow \text{C} + \text{OH}
\end{equation}
which has an activation energy corresponding to 2500~K.
The other significant reactions involve the hydrogenation of CO, resulting in the formation of HCO:
\begin{equation}
\label{eqn:hydrogenation1}
\text{CO} +\text{H} + \text{M} \rightarrow \text{HCO} + \text{M}
\end{equation}
where M designates a third body, or 
\begin{equation}
\label{eqn:hydrogenation2}
\text{CO} +\text{H}_2 + \text{M} \rightarrow \text{HCO} + \text{H.}
\end{equation}
Finally
\begin{equation}
\label{eqn:co2}
\text{CO} +\text{OH} + \rightarrow \text{CO}_2 + \text{H}
\end{equation}
becomes important if the H$_{2}$ abundance is not sufficiently high for hydrogenation.

The reaction rate coefficients, $k$, of the processes described by Equations~\ref{eqn:co_formation} to~\ref{eqn:co2} depend on temperature, $T$, via the modified Arrhenius equation \citep{arrenius}
\begin{equation}
\label{eqn:arrenius}
k(T) = \alpha T ^\beta \exp\left[\dfrac{-E_{a}}{RT}\right]~\text{cm}^3~\text{s}^{-1}\text{,}
\end{equation}
where $\alpha$ and $\beta$ are constants, $R$ is the gas constant, and $E_{a}$ is the activation energy.
With $k(T)$ for radiative association remaining approximately constant around $10^{-16}$~cm$^3$~s$^{-1}$, and the dominant method of destruction (Eqn.~\ref{eqn:h_co}) ranging between $\sim10^{-17}$~cm$^3$~s$^{-1}$ at the coolest Cepheid temperatures and $\sim10^{-15}$~cm$^3$~s$^{-1}$ for the hottest \citep{2015ApJS..217...20W}. 

Each of the reactions in Equations~\ref{eqn:co_formation} through~\ref{eqn:co2} results in a change in the amount of CO present in the Cepheid's atmosphere, hence a change in opacity of the CO band--head.  Figure~\ref{fig:annotated_lightcurve} shows the cycle of destruction and formation of CO molecules in the atmosphere of HV~900 --- an LMC Cepheid with a 47.515~day period. As the Cepheid radius increases, its temperature decreases (with a slight phase lag), allowing the C and O atoms in the Cepheid's atmosphere to once again form CO. This increase in the amount of CO in the Cepheid's atmosphere increases the opacity of the star at 4.6~$\mu$m and the surrounding wavelengths, causing a decrease in the emitted 4.5~$\mu$m flux. This causes a flattening in the [4.5] light curve and induces a variation in the $([3.6]-[4.5])$ colour of the Cepheid. As the star reaches maximum radius, changing from expansion to contraction, it begins to heat up again. The increase in temperature causes the CO molecules to be chemically destroyed once more, allowing the 4.5~$\mu$m radiation to once again escape from the star. This increase in flux boosts the [4.5] magnitude and once again varies the $([3.6]-[4.5])$ colour.

Plateaus are seen in the mid--IR colour curves for Cepheids when the maximum temperature is reached. For the hottest Cepheid temperatures (approximately 6000~K at mean light, see models discussed in Section~\ref{sec:pc_relation}) destruction of CO is the dominant reaction. At this point, the variation in $([3.6]-[4.5])$ colour saturates, and we observe the flat--topped colour curves seen in Fig.~\ref{fig:annotated_lightcurve}. 

\subsection{Cepheid Near--IR and Mid--IR Colour Curves}
\label{sec:is_3p6_ok}
\begin{figure}
\begin{center}
\includegraphics[width=80mm]{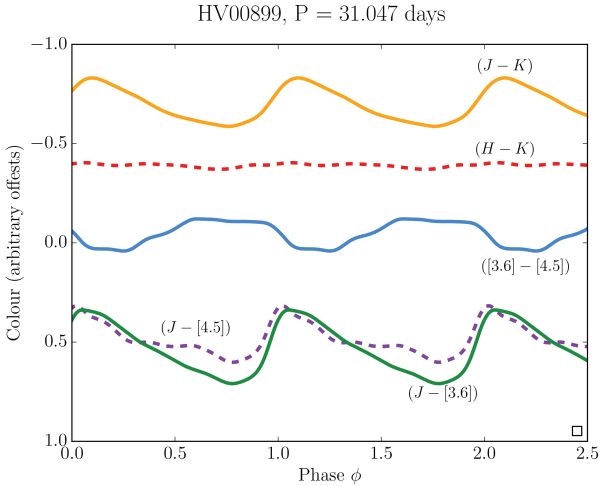}
\caption{A selection of near-- and mid--IR combinations of colour curves for the LMC Cepheid HV~889. The lines are {\sc gloess} fits to data from \citet{2004AJ....128.2239P} for the near--IR and \citetalias{2011ApJ...743...76S} for the mid--IR. The colour curves are phased such that $\phi=0$ is at minimum light in the [3.6] light curve. }
%code single_cepheid_many_colours.py
\label{fig:all_band_col_curves}
\end{center}
\end{figure}

If the 3.6~$\mu$m flux were also affected, either by CO or some other element or molecule, the mid--IR bands would be unsuitable for Cepheid distance measurements. Fortunately, this is demonstrably not the case. The near-- and mid--infrared colour curves in Figure~\ref{fig:all_band_col_curves} illustrate empirically that the CO effect is confined to the [4.5] magnitudes alone, consistent with predictions from model atmospheres \citep{2003IAUS..210P.A20C} and template light curves \citep{2012ApJ...748..107P}.
%Is the 3.6~$\mu$m flux also affected, either by CO or some other element or molecule? Fortunately, this is demonstrably not the case.

\citet{2012ApJ...748..107P} created template Cepheid light curves in 29 photometric bands between 0.3 and 8.0~$\mu$m, using 287 MW, LMC, and SMC Cepheids with $P>10$~days. They define the effect of temperature at a given wavelength as 
\begin{equation}\label{eqn:beta}
\beta_{i} = \frac{\partial \log F_{i}} {\partial \log T}\bigg|_{T_0}
\end{equation}
where $F_{i}$ is the flux in filter $i$ . In this analysis, we adopt the $\beta_{i}$ values given in their table~3. We can quantify the expected temperature effects on a given colour ($X-Y)$ as
\begin{equation}\label{eqn:beta_col}
\Delta\beta_{x, y} = \beta_{x} - \beta_{y}
\end{equation}

The colour curves presented in Figure~\ref{fig:all_band_col_curves} are \textit{observed} colour curves, fit using {\sc gloess}. The original data are from \citet{2004AJ....128.2239P} and \citetalias{2011ApJ...743...76S}.
First we consider the near--IR colour curves involving $J$, $H$ and $K$ only (the top two curves on the plot). The $(J-K)$ curve (orange solid line) shows some temperature variation that comes from the $J$ band (the shortest wavelength, where temperature has the most significant effect, $\beta_J = 2.246 \pm 0.005$, $\beta_K = 1.553 \pm 0.005$, $\Delta\beta_{J, K} = 0.693 \pm 0.007$). As we expect, the $(H-K)$ colour curve (red dashed line), however, is almost completely flat. This is because we have now entered the region where spectra of objects of different temperatures are parallel ($\beta_H = 1.724 \pm 0.005$, $\Delta\beta_{H, K} = 0.171\pm 0.007$ hence $\Delta\beta_{H, K} < \Delta\beta_{J, K}$), so temperature changes result in a much smaller colour change. %These curves look as anticipated. 

The blue, solid curve close to zero (centre of the plot) is the mid--IR $([3.6]-[4.5])$ colour %curve. The variation of this curve us the opposite to the $(J-K)$ curves; i.e. as the $(J-K)$
curve. The variation of this curve is the opposite form to the $(J-K)$ curves; i.e. as the $(J-K)$ colour gets redder, the $([3.6]-[4.5])$ colour gets bluer. This is consistent with the colour change in the IRAC bands being driven by the CO band--head at 4.6~$\mu$m, which theory tells us will show up as a temperature--opposite feature ($\beta_{[3.6]} = 1.442\pm 0.023$, $\beta_{[4.5]} = 1.726\pm 0.020$, hence $\Delta \beta = -0.284 \pm 0.030 $). However, this curve is insufficient to determine if the effect is limited to one filter or affects both.
%alone cannot tell us whether the effect is in both filters or just one.

Next, we consider the colours involving the near--IR bands and [3.6]. For $(J-[3.6])$ (green solid line, bottom of plot) we see that the colour curve has the same shape as the temperature induced near--IR curve.  As expected, it is not flat. At 3.6~$\mu$m two stars of different
temperatures will have parallel spectral slopes, but at the wavelength of the $J$ band they will not, inducing a colour curve in $(J-[3.6])$. It has a slightly higher amplitude than the $(J-K)$ curve. The fact that this curve follows the shape of the pure near--IR colour curves is the first suggestion that CO is not affecting [3.6].

Finally, we examine the  $(J-[4.5])$ colour curve (purple dashed line, bottom of plot). This curve follows the same temperature variation that the $(J-[3.6])$ colour follows, but the amplitude is decreased. This behaviour indicates that the temperature sensitivity in the $J$ band is being directly opposed by the inverse--temperature effect in the [4.5] band, and confirms that the CO effect on Cepheid magnitudes is confined to the [4.5] mid--IR band alone.

\begin{figure}
\begin{center}
\includegraphics[width=80mm]{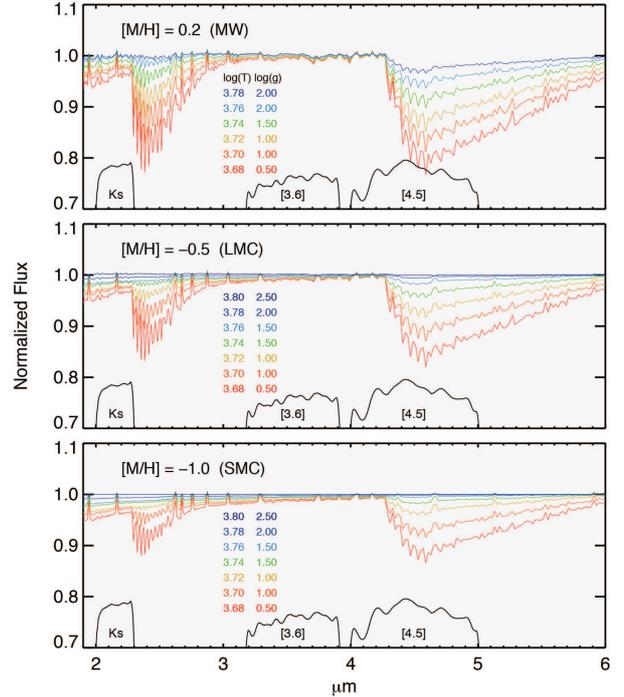} 
\caption{ATLAS9 LTE synthetic spectra for the range of $T_{\text{eff}}$ experienced by Cepheids at MW, LMC and SMC metallicities. Solid black lines represent the near and mid--IR filter response curves. The two CO band--heads at 2.3 and 4.6~$\mu$m can clearly be seen. The 2.3~$\mu$m overtone band--head falls close to the $K_S$ band and does not affect the mid--IR observations. The 4.6~$\mu$m band--head begins in the [4.5] IRAC filer and spreads to longer wavelengths, having no affect on the [3.6] observations. The depth of the CO features increase with decreasing temperature, suppressing the flux in the [4.5] filter at longer periods. The depth of the features also decreases when moving from higher to lower metallicity. These effects are echoed in the colour curves; the amplitude of the colour curves increases with increasing period and metallicity, as is seen in Figure~\ref{fig:example_lcs}.}
\label{fig:temp_spec}
\end{center}
\end{figure}

Figure~\ref{fig:temp_spec} demonstrates how the depth of the CO feature at 4.6~$\mu$m changes with temperature. Each panel shows synthetic spectra from the ATLAS9 models \citep{2003IAUS..210P.A20C} at metallicities corresponding to the MW, LMC and SMC, with typical temperatures of Cepheids with periods between $\sim 6$ and 100~days, assuming the period--temperature relation of \citet{2004A&A...424...43S}. As the temperature decreases, the depth of the CO feature increases. This makes the Cepheid appear bluer in its $[3.6]-[4.5]$ colour. The temperature dependence of the depth of the CO feature is reflected in the Cepheid colour curves as a change in amplitude with period, such that the cooler, long period Cepheids have larger amplitude colour curves. This is seen in the colour curves in Figure~\ref{fig:example_lcs}. Also of note is the change in the depth of the feature with metallicity. This is discussed in detail in Section~\ref{sec:metallicity_effect}. 

These models also demonstrate the utility of the 3.6~$\mu$m observations for precision distance measurements. It is clear that the CO feature falls exclusively within the IRAC [4.5] filter and has no effect on the [3.6] magnitudes. Similarly, the overtone feature at 2.3~$\mu$m, which appears at slightly longer wavelengths than the $K_S$ filter has no effect on the [3.6] band. The colour curves shown in Figure~\ref{fig:all_band_col_curves}, are consistent with the models presented in Figure~\ref{fig:temp_spec}, and we conclude that the IRAC [3.6] band provides an excellent tool for precision distance measurements.

\section{The Period--Colour Relation}
\label{sec:pc_relation}
\begin{figure}
\includegraphics[width=80mm]{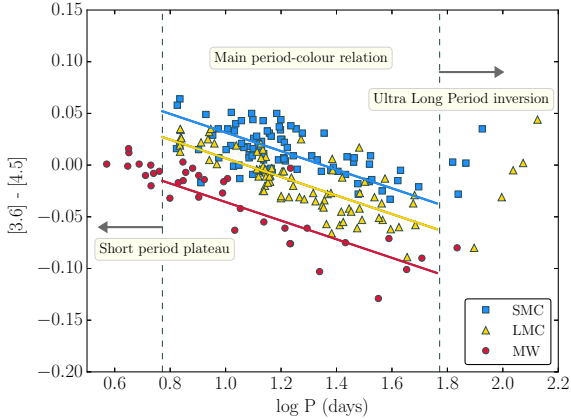}
% code all_gals_pc_relation.py
\caption{Period--colour relation for all three galaxies. The short period plateau is seen in the MW sample, and the long period inversion is in the Magellanic Cloud samples. The mean period--colour relations of each sample are clearly offset from one another, with the offsets moving to {\it redder} colours with {\it decreasing} metallicity.}
\label{fig:pc_allgals}
\end{figure}

\begin{figure}
\begin{center}
\includegraphics[width=80mm]{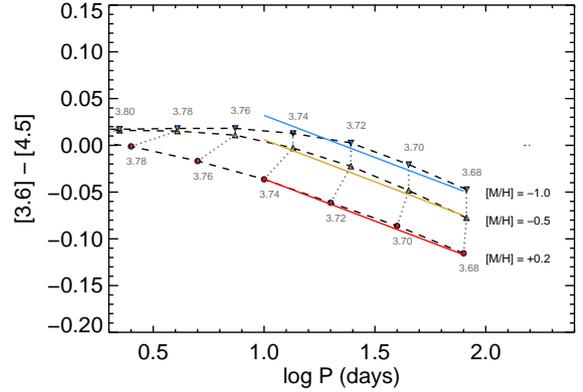} 
\caption{Model PC relations for Cepheids in the MW, LMC and SMC derived by combining the ATLAS9 LTE models from \citet{2003IAUS..210P.A20C} with the $\log P - \log T_{\text{eff}}$ and $\log P - \log g$ relations from \citet{2004A&A...424...43S}. Dashed black lines show the model PC relations corresponding to metallicities representing the MW (bottom), LMC (middle) and SMC (top). Grey dotted lines connecting the points show lines of constant $\log T_{\text{eff}}$. The red, yellow and blue lines show the empirical PC fits to the MW, LMC and SMC presented in Table~\ref{tab:pc_relations_lmc_fit}.}
\label{fig:theory_pc}
\end{center}
\end{figure}

Given the temperature difference between the longer and shorter period Cepheids, and the destruction and formation of CO molecules in a Cepheid's atmosphere is strongly temperature dependent, we should expect to see a mid--infrared Cepheid period--colour relation. Figure~\ref{fig:pc_allgals} shows the period--colour relations for the Cepheids in the three galaxies. Figure~\ref{fig:theory_pc} shows model period--colour relations derived from the ATLAS9 LTE models from \citet{2003IAUS..210P.A20C} combined with the $\log P - \log T_{\text{eff}}$ and $\log P - \log g$ relations from \citet{2004A&A...424...43S}, fitting to their figure 3 for $\log T_{\text{eff}}$, and adopting their equation 49 for $\log g$. We adopt the LMC $\log P - \log T_{\text{eff}}$ relation for the SMC as no suitable relation was found at this metallicity. The three model metallicities ([M/H] = $+0.2$, $-0.5$, $-1.0$) were selected as the closest match to the average metallicities of the Cepheids in each of our samples. 

The Cepheid period--colour relation can be described by several features; the main period--colour relation, the short--period plateau ($P < 6$~d or $\log P < 0.8$, seen mainly in the Galactic Cepheids due to the period distributions of our samples), and the long--period inversion ($P >60$~d or $\log P > 1.8$). Each of these features will now be addressed in turn.

\subsection{The Mid--Infrared Period--Colour Relation Between 6 and 60 Days}
\label{sec:main_midir_pc}
Between approximately 6 and 60 days ($0.8 \leq \log P \leq 1.8$), all three galaxies exhibit well defined period--colour relations. We performed independent unweighted, least--squares fits to each sample for PC relations of the form 
\begin{equation}
\label{eqn:pc_2fit_form}
([3.6] - [4.5]) = a (\log P - 1.0)+ b\text{.}
\end{equation}
The values of $a$, $b$ and their respective uncertainties for each galaxy are given in Table~\ref{tab:pc_relations}. The slopes are in good agreement for each of the galaxies, but with significantly different intercepts. 

\begin{table}
	\centering
	\caption{Coefficients for initial PC relation fits to Equation~\ref{eqn:pc_2fit_form} for each galaxy.} 
	\label{tab:pc_relations}
	\begin{tabular}{l|c|c|c|r} 
		\hline \hline
		Galaxy& \multicolumn{2}{|c|}{Slope} & \multicolumn{2}{|c|}{Intercept} \\
				&  \multicolumn{1}{|c|}{$a$} & \multicolumn{1}{|c|}{$\sigma_{a}$} & \multicolumn{1}{|c|}{$b$} & \multicolumn{1}{|c|}{$\sigma_{b}$} \\

		\hline
		MW & $-0.10$ & $0.03$ & $-0.040$ & $0.010$\\
		LMC & $-0.09$ & $0.01$ & $+0.006$ & $0.004$\\
		SMC & $-0.07$ & $0.01$ & $+0.028$ & $0.003$\\
		\hline
	\end{tabular}
\end{table}

For the remaining analysis, we adopt the slope defined by the LMC sample as it is the most well-defined sample for this period range, and we refit the PC relations using this fixed slope of $-0.09 \pm 0.01$~mag.

The period--colour relations for each galaxy take the form:
\begin{equation}\label{eqn:pc_relation_lmc_slope}
([3.6]-[4.5]) = -0.09 (\log P - 1.0) + c
\end{equation}
The value of $c$ and its uncertainty for each galaxy are given in Table~\ref{tab:pc_relations_lmc_fit}. 

The PC relations from Equation~\ref{eqn:pc_relation_lmc_slope} and Table~\ref{tab:pc_relations_lmc_fit} are plotted in Figures~\ref{fig:pc_allgals} and \ref{fig:theory_pc} as solid coloured lines. It is clear in Figure~\ref{fig:theory_pc} that the empirical PC relations are well described by the model relations in the intermediate period range ($10 < P < 60$~days). Below 10 days the model PC relations begin to plateau for all samples (as described in Section~\ref{sec:short_period}), but there are insufficient data in our LMC and SMC samples to study this effect in detail between 6 and 10 days. We therefore assume a linear relation between $6 < P < 60$ days, consistent with other CCHP works.
\begin{table}
	\centering
	\caption{PC zero--points of Equation~\ref{eqn:pc_relation_lmc_slope} using fixed slope of $-0.09$ defined from LMC Cepheid sample.} 
	\label{tab:pc_relations_lmc_fit}
	\begin{tabular}{l|c|r} % four columns, alignment for each
		\hline \hline
		Galaxy& \multicolumn{2}{|c|}{ Intercept} \\
			&  \multicolumn{1}{|c|}{$c$} & \multicolumn{1}{|c|}{$\sigma_{c}$} \\
		\hline
		MW & $-0.036$ & $0.004$ \\
		LMC & $+0.006$ & $0.002$ \\
		SMC & $+0.032$ & $0.002$ \\
		\hline
	\end{tabular}
\end{table}

%% This subsection needs work. Come back to it %%
\subsection{The Short--Period Plateau}
\label{sec:short_period}
The short period MW sample in Figure~\ref{fig:pc_allgals} demonstrate the high temperature effect that is also seen throughout the Cepheid's pulsation cycle. At periods below $\sim$10~days \citep[$\sim\log P = 1$,  $\log T_{\text{eff}}=3.78$~K,][]{2004A&A...424...43S}, the period--colour relation appears to flatten out, in the same way that the mid--IR light curves plateaued at high temperatures. This is the same mechanism. 

At %these periods, the temperatures of the Cepheids are such that the CO molecules are permanently
these periods, the temperatures of the Cepheids are such that rate coefficient of CO destruction is an order of magnitude higher than for formation \citep{2015ApJS..217...20W}. The CO molecules are chemically dissociated, and the 4.5~$\mu$m flux is not suppressed. The CO band--head has no effect on Cepheids at these periods.

This is also seen in Figure~\ref{fig:theory_pc}, which shows model PC relations derived at metallicities corresponding to the three galaxies. At short periods, the three PC relations converge to a single plateau. This is further demonstrated in Figure~\ref{fig:temp_spec}, where at the highest temperatures temperatures the CO feature is indistinguishable from the continuum. We see the plateau in our MW data, but unfortunately do not have enough Cepheids in the LMC and SMC samples at short period to see the effect at these metallicities. 

\subsection{The Ultra Long--Period Inversion and the Effects of Rotation}
\label{sec:long_period}
%% Sandage + Tamman (1968) -- Fig. 1 shows dropoff of long P ceps
%% Madore + Freedman (1991) -- 'Cepheids with log P > 1.8 are excluded from the least squares fits due to uncertainties in their reddenings and their evolutionary status.'
%% Bird+ (2009) shows separate PL for ULP Ceps.
Ultra long period (ULP) Cepheids (those with periods greater than approximately 60 days) are infamously not well understood \citep{1968ApJ...151..531S, 1991PASP..103..933M}. They are generally excluded from
%They are often excluded from
distance scale works as they tend to fall well below an extrapolation of the mean period--luminosity (PL) relation for shorter periods, and may even adhere to their own distinct PL relation \citep{2009ApJ...695..874B}. The deviation of ULP Cepheids from the mid--infrared PL relation has also been noted in previous CCHP works \citep[e.g.][]{2011ApJ...743...76S}. There have been suggestions that the flattening of the PL beyond 80 -- 100 days may itself be useful as a standard candle \citep[e.g.][and references therein]{2010ApJ...711..808F, 2013IAUS..289..282F, 2011EAS....45..267C}, but currently the ULP PL relation remains uncalibrated.

As can be seen in Figure~\ref{fig:pc_allgals}, the inversion of the period--colour relation at periods greater than 60~days is dramatic. 
This behaviour may give us important clues about the evolution of Cepheids. \citet{2014A&A...564A.100A} demonstrated that the ratio of the rotation rate of a Cepheid to the critical rotation rate, $\omega$, which decreases with increasing mass, is correlated with luminosity, $L$, such that a lower $\omega$ leads to a lower $L$. This decrease in rotation leads to decreased rotational mixing on the main sequence. For the high-mass stars, modification of the CNO abundances by rotational mixing dominates over abundance changes from first dredge--up. In the long-period region the low temperature boosting of CO abundance is counteracted by the lack of CNO enhancement due to diminished rotational mixing. In the long period region of Figure~\ref{fig:pc_allgals} (i.e. $1.8 \leq \log P \leq 2.2$), the Cepheids have approximately the same $T_{\text{eff}}$  \citep[as derived from their $(V-K)$ colours;][]{1971MNRAS.152..121P, 1999AJ....117..521V}, so the change in %mid--IR colour must be driven by the changes in CNO surface abundances and $\log g$ driven by
mid--IR colour may plausibly be driven by the changes in CNO surface abundances and $\log g$ driven by decreases in $\omega$ with increasing mass. Exacerbating this effect could be the lack of first dredge up for high mass Cepheids, as proposed by \citet{2000ApJ...543..955B}, further diminishing the surface abundances of CNO. A full investigation of the ultra long period inversion is beyond the scope of this paper.

\section{The Effect of Metallicity}
\label{sec:metallicity_effect}
Figure~\ref{fig:pc_allgals} shows the PC relations for all three galaxies plotted on the same scale. The three PC relations follow the same slope, and are systematically offset by approximately the same amount. The LMC has an average metallicity of approximately half solar, and the SMC has an average metallicity of approximately one quarter solar \citep[][using a combination of H{\sc ii} regions and F supergiants]{1992ApJ...384..508R}. Figures~\ref{fig:pc_allgals} and \ref{fig:theory_pc} suggest that metallicity is the reason for the offset zero--points. 

The simulated spectra summarised in Figures~\ref{fig:temp_spec} and \ref{fig:theory_pc} add further weight to this conclusion. Figure~\ref{fig:temp_spec} shows how the depth of the CO feature, which corresponds to the amplitude of the colour effect, changes with metallicity. The empirical PC relations are well matched by simulated PC relations corresponding to the average metallicities of the three galaxies in Figure~\ref{fig:theory_pc}. This is further evidence that the offset
%This confirms that the offset
mid--IR PC relations  are due to differences in the average metallicity of the host populations. 

Rather than consider the galaxies as mono--abundance populations, we can use the metallicities compiled by \citetalias{2014A&A...566A..37G} to test metallicity directly for individual Cepheids. In order to ensure the best quality calibration of the metallicity effect we only use those stars which have full phase light curves from {\it Spitzer}. The full \citetalias{2014A&A...566A..37G} sample has a much larger number of MW Cepheids available, but these do not have full phase coverage by {\it Spitzer} at this time. Spectroscopic metallicities are available for 33 MW, 20 LMC and 13 SMC Cepheids in common with our {\it Spitzer} sample. Constraining the sample to Cepheids with periods between 6 and 60 days 
%(the region where we believe the PC relation is linear),
(the region where the PC relation is empirically consistent with linear),
our sample reduces to 24, 18 and 10 in the MW, LMC and SMC, still with enough overlap to quantify the effect of metallicity on the mid--infrared colour of Cepheids.

\begin{figure}
\includegraphics[width=80mm]{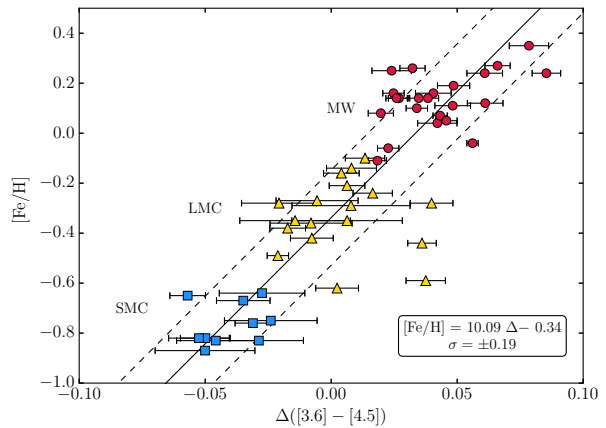}
\caption{Effect of metallicity on the Cepheid mid--IR colour. The LMC period--colour relation has been chosen as fiducial, and $\Delta([3.6]-[4.5])$ indicates the deviation in colour from that relation at a given period. The [Fe/H] values are the spectroscopic values for Cepheids in common with \citetalias{2014A&A...566A..37G}. There is a clear trend of mid--infrared colour with [Fe/H], demonstrating that the IRAC colour can be used as a reliable metallicity indicator for Cepheids in nearby galaxies. Y Oph was removed from the MW sample as its large deviation from the mean PC relation cannot be explained. Removing this star has a negligible effect on the result. }
\label{fig:col_metals}
\end{figure}

\begin{figure}
\includegraphics[width=80mm]{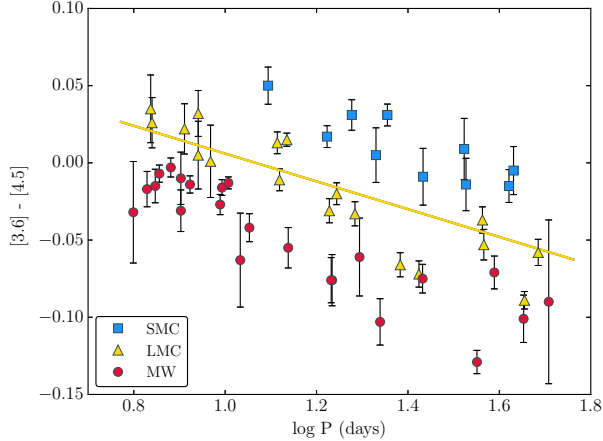}
\caption{Period--colour relation for Cepheids with with metallicities used in Figure~\ref{fig:col_metals} to derive the colour--metallicity relation. The solid yellow line represents the fiducial period--colour relation derived from the full sample of LMC Cepheids.  }
\label{fig:metals_pc}
\end{figure}

In Figure~\ref{fig:col_metals}, the LMC PC relation is selected as fiducial. $\Delta([3.6]-[4.5])$ is defined as the deviation of a point from the mean LMC PC relation. The [Fe/H] abundances are taken directly from \citetalias{2014A&A...566A..37G}. As Figure~\ref{fig:col_metals} shows, there is a clear trend of colour difference with metallicity, demonstrating that the different zero--points seen in Fig.~\ref{fig:pc_allgals} are due to the different average [Fe/H] abundances of the host galaxies. Figure~\ref{fig:metals_pc} further emphasises this point, showing now the period--colour relation for only those stars with spectroscopic metallicities. Here it can clearly be seen that the three samples are following offset period--colour relations, following the trend set down by their average metallicities. 

\section{Mid--IR Colour as a Metallicity Indicator}
\label{sec:metallicity_indicator}

Figures~\ref{fig:pc_allgals} and \ref{fig:col_metals} demonstrate the theoretical and empirical utility of the mid--infrared period--colour relations  as a metallicity indicator. For populations of Cepheids, the average metallicity can be estimated by determining the zero--point offset of the PC relation from the LMC. However, the true power of the mid--infrared colour as a metallicity indicator is for the individual Cepheids. Very few spectroscopic measurements of the metallicities of LMC and SMC (or more distant) Cepheids currently exist, and they are time consuming to obtain. Using the data shown in Fig.~\ref{fig:col_metals}, the metallicity of an {\it individual} Cepheid can be obtained using the relation 
\begin{equation}\label{eqn:ind_cep_metal}
\text{[Fe/H]} = 10.09\Delta([3.6]-[4.5]) - 0.34\text{,}
\end{equation}
where 
\begin{equation}\label{eqn:delta}
	\Delta([3.6]-[4.5]) = ([3.6]-[4.5])_{\text{LMC}} - ([3.6]-[4.5])\text{,}
\end{equation}
$([3.6]-[4.5])_{\text{LMC}}$ is the mid--IR colour predicted from the fiducial LMC PC relation defined by Equation~\ref{eqn:pc_relation_lmc_slope} and Table~\ref{tab:pc_relations}, and $([3.6]-[4.5])$ is the measured mid--IR colour of the Cepheid. This empirical relationship has a standard deviation of 0.20~dex. The uncertainties on the MW [Fe/H] abundances range from 0.05~dex to over 0.25~dex per Cepheid, with the Magellanic Cloud Cepheids estimated to be subject to an internal uncertainty of $\sim 0.10$~dex \citep{2008A&A...488..731R}. 

This relation is defined in terms of the {\it Spitzer} IRAC mid--infrared bands. However, 2018 will see the launch of {\it JWST}, which will supersede {\it Spitzer} in terms of mirror size and resolution. {\it JWST} has the potential to observe many thousands of Cepheids in the MW, Magellanic Clouds, Local Group and beyond using NIRCam \citep{2005SPIE.5904...21B}. The NIRCam F356W and F444W bands are extremely similar to those of IRAC \citep{2006SSRv..123..485G}, meaning that the CO band--head that is fundamental to this technique will still be present in the the longer wavelength filter. 
Using the filter curves available from STScI in November 2015, we recreated the synthetic PC relations in the {\it JWST} NIRCam bands at the average MW, LMC and MW metallicities. We can predict the NIRCam equivalent of Equation~\ref{eqn:ind_cep_metal}:
%-10.3031411425 -0.188971357893
\begin{equation}\label{eqn:ind_cep_jwst}
	\text{[Fe/H]} = 10.30 \Delta([F356W]-[F444W]) - 0.19\text{.}
\end{equation}
%Using the Using the 
Using the {\it JWST}/NIRCam prototype exposure time calculator\footnote{\url{http://jwstetc.stsci.edu/etc/}} we predict that we will be able to measure the metallicities of individual Cepheids to the same accuracy as presented in this work, in galaxies as distant as NGC~4258 \citep[7.6~Mpc,][]{2013ApJ...775...13H}. This will be a huge improvement over what is currently achievable with {\it Spitzer}. We will be able to map out the whole Local Group in the three spatial dimensions using the F356W band, and add the metallicity dimension using the mid--IR colour. NGC~4258 is a key galaxy for the Cepheid distance scale as it is home to a megamaser system which can be used to calculate an accurate and precise geometric distance. This galaxy not only provides an independent cross check of the Cepheid calibration, but is beyond the Local Group. This large distance places it in an environment more comparable to supernova host galaxies, where dynamical effects are less significant. A detailed understanding of NGC~4258, including a study of its metallicity structure, will be important for both Cepheid studies and cosmology.

\section{Conclusions}
\label{sec:conclusions}
The cyclical colour variation and period--mean--colour relation seen in mid--IR observations of Cepheids is due to the CO band--head at 4.6~$\mu$m, which is aligned with the IRAC [4.5] filter. The variation in colour of a single Cepheid is due to the CO being chemically destroyed at high temperatures, and recombining at lower temperatures. The same effect is seen in the mean--PC relation, where the period of a Cepheid in the sample is intrinsically linked to its mean temperature. 

We have discovered that the mid--IR period--colour relation for Cepheids with $6 \leq P \leq 60$~d is dependent on metallicity. We have calibrated this effect, such that the metallicity of an individual Cepheid in this period range can be predicted via the relation presented in Eqn.~\ref{eqn:ind_cep_metal}, which has a dispersion of 0.20~dex. This is competitive with the uncertainties on individual Cepheid metallicities in the MW, LMC and SMC. With the advent of {\it JWST}, we will be able to make mid--IR photometric metallicity measurements of individual Cepheids of this quality out to the distance of NGC~4258 \citep[$d = 7.6$~Mpc,][]{2013ApJ...775...13H} with ease.

The inversion of the PC relation at long periods is consistent with a reduction in rotational mixing in high mass Cepheids. This is consistent with the predictions of \citet{2014A&A...564A.100A}, and provides further evidence for rotation as a solution to the Cepheid mass discrepancy, rather than convective core overshooting. 

Through the study of Cepheid light curves and synthetic spectra we have confirmed that the temperature dependent effect is confined to the [4.5] {\it Spitzer} band, and does not contaminate the [3.6] band. We conclude that the 3.6~$\mu$m period--luminosity relation is suitable for measuring high precision Cepheid distances.

\section*{Acknowledgements}
\label{sec:acknowledgements}
We thank Eric Persson for the many fruitful discussions that made this work possible, and Meredith Durbin for her comments on this paper. We thank Giuseppe Bono and Richard Anderson for insightful discussions regarding Cepheid pulsation theory and the effect of rotation. We also thank the anonymous referee for their comments and advice.

This work is based on observations made with the Spitzer Space Telescope, which is operated by the Jet Propulsion Laboratory, California Institute of Technology under a contract with NASA. Support for this work was provided by NASA through an award issued by JPL/Caltech. 

This research has made use of the NASA/IPAC Extragalactic Database (NED) which is operated by the Jet Propulsion Laboratory, California Institute of Technology, under contract with the National Aeronautics and Space Administration.  We acknowledge the use of NASA's \textit{SkyView} facility (http://skyview.gsfc.nasa.gov) located at NASA Goddard Space Flight Center.

%%%%%%%%%%%%%%%%%%%%%%%%%%%%%%%%%%%%%%%%%%%%%%%%%%



\bibliographystyle{mnras}
\bibliography{metallicity_2015_mnras}


%%%%%%%%%%%%%%%%%%%%%%%%%%%%%%%%%%%%%%%%%%%%%%%%%%

%%%%%%%%%%%%%%%%% APPENDICES %%%%%%%%%%%%%%%%%%%%%

%\appendix



%%%%%%%%%%%%%%%%%%%%%%%%%%%%%%%%%%%%%%%%%%%%%%%%%%

% Don't change these lines
\bsp	% typesetting comment
\label{lastpage}
\end{document}